# Energy Sensing and Monitoring Framework with an Integrated Communication Backbone in Energy Efficient Intelligent Buildings

Jianli Pan[1, 3, a], Shanzhi Chen[2, b], Raj Jain[3, c], Subharthi Paul[3, d]

[1]Beijing University of Posts and Telecommunications, Beijing, China

[2]State Key Laboratory of Wireless Mobile Communications, China Academy of Telecommunications Technology

[3]Dept. of Computer Science and Engineering, Washington University in Saint Louis, USA

[a]pann1588@gmail.com, [b]chenshanzhi@yahoo.com.cn, [c]jain@cse.wustl.edu, [d]pauls@cse.wustl.edu

**Keywords:** Sensing, Monitoring, Energy Efficiency, Intelligent Building, Communication, Sustainability, Smart Energy, Smart Box, Cloud Computing, Smart Phone.

**Abstract.** Building environments are significant sources of global energy consumption. To create energy efficient buildings, the first step is to sense and monitor all the energy-consuming appliances in the buildings and record all the energy consumption information. After that, appropriate energy saving policies can be decided and the instructions can be sent to the control devices to apply the energy saving adjustments. To do that, in-building two-way communication networks are needed to connect all the sensors to collect information as well as to send control instructions. However, most of the current devices are provided by separate manufacturers and with separate network infrastructures and so there is not much integration and interaction among different subsystems. In this paper, we envision a new energy sensing and monitoring framework with integrated communication backbone in the intelligent building environments. Specifically, through comprehensive comparisons and investigations, we study different candidate communicating media and protocols like wireline, wireless, and power-line communications technologies that potentially can be used in the intelligent buildings to realize the goals of coordination, integration, and energy efficiency. Also, we propose an extension "smart box" for integration of the devices before the maturity of the standardization process. Cloud computing and smart phone technologies are also introduced to realize the goals of improving energy efficiency and promote global sustainability.

## 1. Introduction

It is known that most of the people stay in different building environments most of the time. Buildings have become very significant energy consumption sources in all countries. Taking United States as a typical example, buildings consume about 40% of the total energy and 70% of electricity in total [1]. Also, about 38% of the carbon dioxide emissions come from buildings. Given these facts, it is very important to improve the building environments for occupants' comfort and energy efficiency. It also has great impacts on the occupants' productivity as well as the global sustainability. To realize such goals, the first thing that need to be done is to deploy distributed real-time sensing and monitoring systems inside the buildings to know every aspect of the building status before any further improvements are made. Thus, in-building two-way communication networks are needed to connect all the sensors to collect information as well as to send control instructions. However, currently, the subsystems of most of the conventional buildings are provided by separate manufacturers using proprietary protocols and they also use separate communication networks to connect their own devices. It potentially leads to two major problems. The first is that it is very difficult to create any interaction or synergy among these subsystems, or to create an integrated and centralized sensing and monitoring system that can easily get information from these subsystems. Secondly, the construction and maintenance costs of using individual communication networks can be significantly higher than using an individual integrated and consolidated backbone communication network.

Hence, in this paper, from an energy efficiency perspective, we have two research goals. The first is to propose a new energy sensing and monitoring framework with integrated communication backbone in the intelligent building environments. Specifically, through comprehensive comparisons and investigations, we study different candidate communication media and protocols that potentially can be used in the intelligent building environments to realize the goals of coordination, integration, and energy efficiency. We consider a consolidated backbone in-building communication network which takes advantage of the suitable media like wireline, wireless, and power-line communication technologies to realize the goals of energy efficiency in the building environments. Secondly, before standardization process is mature when these devices and subsystems can be naturally integrated and work together, we propose an interim method which provides an extension "smart box" for those devices to integrate and interact. The functions of the conventional devices vary and some may not have any communication or control capabilities. To include such devices into the sensing and monitoring framework, we need extend the simple sensors to enable a specific set of communication and control functions.

The rest of this paper is organized as follows. Section 2 describes the integrated and consolidated in-building communication backbone. Section 3 is about the "smart box" idea for interim usage before the maturity of the standardization process. Section 4 includes some discussions from energy efficiency perspective. In Section 5, we present conclusions.

## 2. Integrated Communication Backbone

Buildings are complex systems. The communication infrastructures in the intelligent building need to accommodate all kinds of applications including in-building facilities control, HVAC (Heating, Ventilation and Air Conditioning), lighting, fire and safety control, and building access control. However, they are always constructed by different providers in separate contracts. Each contract may install communication networks by themselves. This introduces a lot of duplicated construction costs and significant wastes. Also, separate contracts in communication networks also bring additional difficulties and costs in maintaining these communications networks separately. Thus, the benefits of integrating these communication networks are straightforward. We first consider the physical layer media that can potentially be used in creating such integration.

**2.1 Physical Layer Media.** Intuitively, for the building environments, communication technologies based on multiple types of media can be used. Typical examples include wireless radio frequency, Ethernet wire, optical fiber, power line, cable, twisted pairs, etc. Depending on the details, devices using all these physical communication media can all be available in a specific building for different subsystems.

• **Radio Frequency.** Many communication technologies use wireless radio frequency to transmit signals and communicate with each other. IEEE 802.11b based WiFi [2] using radio frequency inside the buildings can be deployed to interconnect the smart sensors that can be accessed by wireless signals. Other wireless sensing and monitoring devices use Bluetooth [3], Ultra-Wide band (UWB) [4], etc. In the building environments, wireless technologies are very suitable for interconnecting all kinds of leaf sensing and monitoring nodes before reaching the communication backbone, especially for those places where the alternatives like wireline are hard to reach or with highly dynamic topology.

• **Wireline.** Wireline consists of many different types. It includes optical fiber, network cable, TV cables, twisted cables, etc. Actually, the original Internet infrastructure was developed primarily from the wireline cables. For building environments, to build a consolidated and integrated backbone communication network, wireline is a good candidate due to its reliability, high bandwidth, and compatibility. The LAN (Local Area Network) [5] technology which was successful in the Internet can be tuned to fit such requirements quickly with relatively low costs. Many protocols across the OSI (Open Systems Interconnection) protocol stack [9] can run over wireline infrastructures.

• **Power line.** Power line communication is relatively inexpensive and there is no need to deploy dedicated expensive routers or switchers to relay the signals. It is widely used for in-home Internet access and applications without wireless signal relaying. However, primary factors limiting power line communication applications are the power signal transmission range issue and the signal interference issue. Also, transformers may prevent the signals being transmitted farther enough. Thus, power line communication is more often used for small-scale in-home building applications. Similar to the wireless technologies, in the general building environments, power line communication technology can be used in leaf sensing and monitoring nodes as a complement before reaching the communication backbone

**2.2 Communication Protocols**. As the most successful data transmission protocols in Internet, TCP/IP (Transmission Control Protocol/Internet Protocol) [10] protocols suite is very open and can run on many types of lower layer media and support many types of upper layer applications. Hence, for intelligent building scenario, it is naturally a good candidate to integrate devices from various hardware and software vendors. On the consolidated communication backbone in the intelligent buildings, it is also desirable that all the applications run TCP/IP directly or adopt designs that are compatible with TCP/IP. It will help the specific application achieve global scale functionalities since the TCP/IP is globally accepted and numerous applications are currently running on top of it globally. We illustrate these concepts in a simple diagram in Fig. 1.

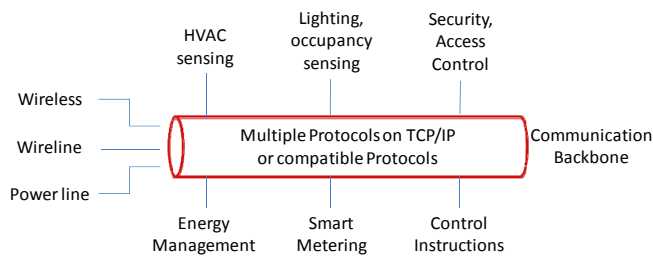

Fig. 1. Multiple applications run on the communication backbone which integrates leaf sensing nodes with multiple physical media

As shown in Fig. 1, multiple intelligent building applications can run on the communication backbone which assembles the leaf sensing nodes' communication using multiple physical media. Some typical compatible protocols for in-building communication and control are BACnet (Building Automation and Control Networks) [6] and LonWorks [7]. They also represent two popular standardization efforts in the intelligent building industry. With the progress of standardization, more and more devices will be integrated into a central sensing, monitoring and control system through the communication backbone. Duplicated construction and maintenance can be avoided and the optimization can be achieved.

## 3. Extending Existing Sensors with Smart Boxes

The conventional buildings are equipped with many different sensors and meters. Some of them are smart or with intelligent communication capability and others are not. So, to create integrated monitoring and control systems in the intelligent buildings, there is a need to extend the functions of these sensors to enable them to send data in standard recognizable formats and receive and execute control instructions from the central servers, especially if they are manufactured by different vendors with different capabilities and features. We propose a "smart box" concept which requires very simple knowledge from the existing sensors and add additional communication and control functions upon the original functions. The extended sensors and meters will function just like the complete new smart sensors and meters with integrated communication and control capabilities. In our proposed idea, the additional functionalities are modularized and only the required ones are deployed depending on the situation. For example, for existing sensors, as long as at least the data can be read out according to the specification of the initial manufacturers, the extension "smart box" can provide the formats transformation and storage capability for the metered and sensed data. After that, the communication module can send the stored data to the central control servers using networking protocols. Besides these modules, there is also a control instruction execution module that executes the received instructions. These modules are especially useful for those smart panels and smart breakers which can control the energy provisioning for a group of appliances. For example, if the central servers realized that the appliances of a specific section grouped under a smart panel are mostly unused, then the control instructions can be sent to the smart breakers or panels to shut off these devices to save energy. An intuitive smart box concept illustration is shown in Fig. 2.

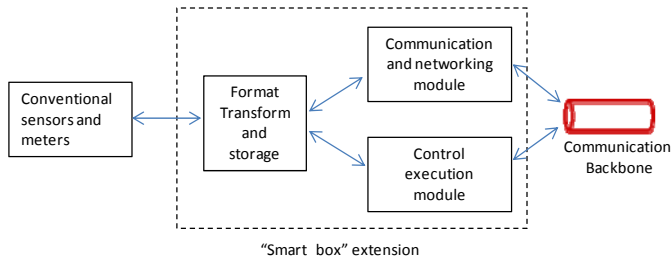

Fig. 2. Smart box extension for traditional sensors and meters

Note that depending on the detailed functions of the traditional sensors, the smart extension modules in the dotted box can vary. In other words, the smart box extension is customizable according to the users' own demands. After associating with the smart boxes, the new sensors and meters can connect to the communication backbone with wireless, wireline, or powerline communication technologies.

## 4. Energy Efficiency Perspective

The contents presented in the above sections are both focused on creating intelligent systems inside the intelligent buildings. Overall, making the buildings more intelligent can bring convenience for the building operation and maintenance, bring comforts to the occupants, and lower the costs for the building stakeholders. It can also make the buildings much more energy efficient and help reduce the overall energy consumption and promote the global sustainability efforts. By connecting the individual intelligent buildings with the smart grid and getting the real-time energy-pricing information, the buildings can avoid peak time energy consumption by the so called demand response [8] technology.

**4.1 Buildings as Both Consumers and Providers.** Future intelligent buildings may also have their own renewable energy generating capabilities by installing solar panels, wind turbines, and biofuel generators. Each building consumes energy and generates energy. So it is efficient to connect these buildings and share energy with each other to maximize the overall efficiency and minimize the overall energy import from the outside grid. To realize such a goal, it is important to create a common "knowledge plane" for a group of geologically closely located intelligent buildings. Such knowledge plane can be created on a central server in a community "microgrid" [9][10]. The central servers keep collecting real-time energy consumption and generation information from multiple buildings, and after modeling and dynamic scheduling using appropriate algorithms, they send out detailed energy allocation and distribution instructions to each intelligent building and apply the detailed strategies. Generalizing such methods into even larger scales can generate much more impact to the global energy efficiency and sustainability.

**4.2 Cloud Computing Applications**. The Internet development is migrating dramatically from the connection-based mode into content-based mode. Such trend is triggered and led by the content services based major companies like Google, Facebook, Apple, etc. The Internet computing capabilities are turned into "utility-like" services which liberate the users from owning and maintaining expensive high-performance computing devices. In intelligent buildings and microgrids cases, cloud computing can be a very useful and convenient tool to allow the buildings to use intelligent Internet access, storage, and computing services flexibly without significant financial burdens. Cloud computing platform also enables fast and easy application development and implementation.

**4.3 Smart Phones Involving Broad Participation**. Studies show that common occupants' awareness and participation can significantly impact the overall energy efficiency in a specific intelligent building [11][12]. When the intelligent building system is combined with the mobile smart phone technologies, it can actively involve every common occupant into the global energy-saving efforts which can potentially make a huge difference. A typical effort is that we can develop an easy-to-use mobile smart phone based application that allows the occupants to monitor and control their own energy consumption. Such methods can bring broad social benefits and attract broad attention from general public hence help promote the global energy-saving efforts and global sustainability.

## 5. Conclusions

In this paper, we envisioned an energy sensing and monitoring framework for the intelligent building environments which utilizes a consolidated and integrated communication backbone. It avoids the current problem in conventional buildings where duplicated and separate communication network deployments in subsystems cause serious costs and maintenance burdens. Moreover, we proposed to extend the conventional sensors with a "smart box" which contains multiple modules that enable the functions like data formats transformation, storage, data transmission and control instructions execution. It can integrate the conventional sensors into the new smart infrastructure even when the standardization is not mature yet. Finally, cloud computing technology is introduced to facilitate larger scale applications, and smart phone technology is used to involve the awareness and participation from the general public.

## Acknowledgements

This work was financially supported by the National High-Technology Program of China (863) project No.2011AA01A101 and International Joint Research project of the Ministry S&T of China No.2010DFB13020.